\documentclass[a4paper,11pt]{article}
\usepackage{pos}
\usepackage{bm}

\renewcommand{\thefootnote}{\fnsymbol{footnote}}

\title{Recent update of nucleon axial-vector charge with the PACS10 superfine lattice}
\author*[a]{Masato~Nagatsuka}
\author[b]{Yasumichi~Aoki}
\author[c]{Ken-Ichi~Ishikawa}
\author[d]{Yoshinobu~Kuramashi}
\author[a]{Shoichi~Sasaki}
\author[e]{Kohei~Sato}
\author[d,f]{Eigo~Shintani}
\author[g]{Ryutaro~Tsuji}
\author[h]{Hiromasa~Watanabe}
\author[i,d]{Takeshi~Yamazaki}
\affiliation[]{\normalsize{\bf \sffamily \hspace{50mm}(PACS Collaboration)}}

\affiliation[a]{Department of Physics, Tohoku University, 980-8578, Sendai, Japan}

\affiliation[b]{RIKEN Center for Computational Science, 650-0047, Kobe, Japan}

\affiliation[c]{
Core of Research for the Energetic Universe,
Graduate School of Advanced Science and Engineering, Hiroshima University 739-8526, Higashi-Hiroshima, Japan}

\affiliation[d]{
Center for Computational Sciences, University of Tsukuba, 305-8577, Tsukuba, Japan}

\affiliation[e]{
Office of the President, Seikei University, 3-3-1 Kichijoji-Kitamachi, Musashino-shi, Tokyo 180-8633, Japan
}

\affiliation[f]{Graduate School of Engineering, University of Tokyo, Hongo 7-3-1, Bunkyo, Tokyo, Japan}

\affiliation[g]{High Energy Accelerator Research Organization (KEK),
  305-0801, Tsukuba, Japan}

\affiliation[h]{
Yukawa Institute for Theoretical Physics, Kyoto University, Kyoto 606-8502, Japan
}

\affiliation[i]{
Institute of pure and Applied Sciences, University of Tsukuba, 305-8571, Tsukuba, Japan
}
\emailAdd{masato.nagatsuka.r4@dc.tohoku.ac.jp}
\abstract{
We update the results of the nucleon axial-vector charge with the third ensemble of the PACS10 gauge configurations, which are generated by the PACS Collaboration at the
physical point with lattice volume larger than $(10\;{\rm fm})^4$ and three
different lattice spacings, 0.085 fm (coarse), 0.063 fm (fine) and 0.041 fm (superfine). 
Although the results of the first two ensembles generated at the coarse and fine lattice spacings
are published, our study using the third one generated at the superfine lattice spacing is still underway. In this work, the low-energy relations arising from the partially conserved axial-vector current (PCAC) relation are also examined in terms of the nucleon three-point functions to verify whether the lattice QCD data correctly reproduces the physics in the continuum within the statistical accuracy.
}

\FullConference{The 42nd International Symposium on Lattice Field Theory (LATTICE2025)\\
2-8 November 2025\\
Tata Institute of Fundamental Research, Mumbai, India\\}

\begin{document}
\maketitle

\section{Introduction}

We have extensively studied the axial structure of the nucleon in lattice QCD at the physical 
point using two sets of the PACS10 gauge configurations in our previous works~\cite{Shintani:2018ozy,Tsuji:2022ric,Tsuji:2023llh}. The axial-vector charge
$g_A$ is especially regarded as the most important benchmark for the nucleon structure, as 
it has been measured precisely by the experiments. In our previous studies, the results of $g_A$
obtained from both the coarse and fine lattices, have achieved that all major sources of systematic uncertainties from the chiral extrapolation, finite size effect, excited-state
contamination and discretization effect are under control by a level of statistical precision of about 2\%.  However, additional lattice simulations using the third PACS10 ensemble are required 
for achieving a comprehensive study of the discretization uncertainties and then taking the continuum limit of our target quantities. 

The PACS10 gauge ensembles are generated by PACS Collaboration using stout-smeared $O(a)$-improved Wilson-clover quark action and Iwasaki gauge action,
and they are characterized by the spatial extent of 10 fm and three lattice spacings 0.085 fm (coarse), 0.063 fm (fine) and, 0.041 fm (superfine)~\cite{Yamazaki:2024otm}.
These ensembles can also reproduce pion masses of 135, 138, and 142 MeV, respectively, which are very close to the physical pion mass.
A brief summary of the simulation parameters is given in Table.~\ref{tab:simulation_details}.
Therefore, these three ensembles allow us to extract physical quantities in the continuum limit at the physical pion mass with negligible finite-volume effects.
The simulations using the coarse and fine ensembles had already been performed~\cite{Shintani:2018ozy,Tsuji:2022ric,Tsuji:2023llh},
and the novel approach for removing $\pi N$-state contamination was also proposed in Ref.~\cite{Aoki:2025taf}. 
The preliminary results obtained using the superfine lattice were reported in Ref.~\cite{Tsuji:2024scy}.

%
% TABLE 1
%
\begin{table*}[b]
\caption{
Parameters for three sets of the PACS10 ensembles.
See Refs.~\cite{Shintani:2018ozy, Tsuji:2022ric, Tsuji:2023llh} for further details.
\label{tab:simulation_details}}
\centering
\begin{tabular}{l|cccccccc}
\hline \hline
 Label & $\beta$ &$L^3\times T$ & $a$ [fm] %$a^{-1}$ [GeV] 
 &  $\kappa_{ud}$ & $\kappa_{s}$ &$c_{\mathrm{SW}}$ &  $m_\pi$ [GeV] \\
          \hline
 PACS10/L256 %(superfine) 
 &2.20& $256^3\times 256$ 
 &0.041 %&4.8
 & 0.1254872 & 0.1249349 & 1.00 & 0.142\\
 PACS10/L160 %(fine)   
 & 2.00& $160^3\times 160$ 
 &0.063 %3.1
 & 0.125814  & 0.124925  & 1.02 & 0.138\\
 PACS10/L128 %(coase) 
 & 1.82& $128^3\times 128$ 
 &0.085% 2.3
 & 0.126177  & 0.124902  & 1.11 & 0.135\\
\hline \hline
\end{tabular}
\end{table*}

In this work, we first present an update to the axial-vector charge calculated at the superfine lattice spacing.
Additionally, an assessment of the discretization uncertainty in our calculations will be presented through a comparison of 
partially conserved axial-vector current (PCAC) 
quark masses defined in different ways based on the axial Ward-Takahashi identity, as described in Ref.~\cite{Tsuji:2025quu}. This consideration exposes 
the potential effect on usage of the local axial-vector current, which does not take into account $O(a)$-improvement of the axial-vector current. 

\section{Method}
In this section, we briefly review the standard plateau method for calculating the matrix elements associated with nucleons.
In order to discuss the size of the discretization uncertainty on the {\it lattice local axial-vector current}, we also introduce two types of PCAC quark mass, which are determined from the two-point functions of the pion and the three-point functions of the nucleon.

First of all, the nucleon (proton) interpolating operator with a three dimensional momentum $\bm{p}$ is given by
\begin{align}
\label{eq:NucOP}
N(t, \bm{p})=\sum_{\bm{x}}e^{-i\bm{p}\cdot\bm{x}}\varepsilon_{abc}\left[
u_a^T(t,\bm{x})C\gamma_5 d_b(t,\bm{x})
\right]u_c(t,\bm{x})
\end{align}
with the charge conjugation matrix, $C=\gamma_4\gamma_2$. The superscript $T$ denotes
a transposition, while the indices $a$, $b$, $c$ and $u$, $d$ label the color and the flavor, respectively. 
The two-point function of the nucleon
carrying the momentum $\bm{p}$ is given by
\begin{align}
C_N(t-t_\mathrm{src};\bm{p})=\frac{1}{4} \mathrm{Tr}\left\{
        \mathcal{P}_{+}
        \langle N(t, \bm{p})
        \overline{N}(t_{\mathrm{src}}, -\bm{p})
        \rangle \right\}
\end{align}
with $\mathcal{P}_{+}=(1+\gamma_4)/2$
which can eliminate the unwanted contributions from the opposite-parity state for $\bm{p}=\bm{0}$~\cite{Sasaki:2001nf}.

In general, when the momentum $\bm{p}$ ($\bm{p}^\prime$) is given to the initial (final) state located at the time slice $t_{\rm src}$ ($t_{\rm snk}$), the isovector axial-vector matrix element of the nucleon can be obtained from the nucleon three-point function
\begin{align}
    \label{eq:three_pt_func}
    C^{5k}_{A_\alpha}
    (t,t_\mathrm{sep}; \bm{p}^{\prime}, \bm{p})
    =
    \frac{1}{4} \mathrm{Tr}
    \left\{
        \mathcal{P}_{5k}\langle 
        N(t_{\mathrm{snk}}, \bm{p}^{\prime})
        A_{\alpha}
        (t; \bm{q})
        \overline{N}(t_{\mathrm{src}}, -\bm{p})
        \rangle
    \right\},
\end{align}
where the isovector axial-vector current $A_{\alpha}$ located at the time slice $t$,
carrying the momentum transfer $\bm{q}=\bm{p}-\bm{p}^\prime$. The projection
operator $\mathcal{P}_{5k} = \mathcal{P}_{+}\gamma_5\gamma_k$ ($k=1,2,3$)
appearing
in Eq.~(\ref{eq:three_pt_func}) means that the $k$ direction is chosen as the polarization direction.
It is important to recall that, under the exact SU(2) isospin symmetry, the  nucleon three-point functions with the isovector currents do not
receive any contributions from the disconnected diagrams.

In a conventional way to extract the nucleon form
factors, the following ratios are constructed by appropriate
combinations of the nucleon two-point and three-point functions with
the source-sink separation ($t_\mathrm{sep}\equiv t_\mathrm{snk}-t_\mathrm{src}$) 
as
\begin{align}
\mathcal{R}^{5k}_{A_\alpha}(t,t_\mathrm{sep}; \bm{q})
=
\frac{C^{5k}_{A_\alpha}(t,t_\mathrm{sep};\bm{p}^\prime,\bm{p})}{C_\mathrm{2}(t,t_\mathrm{sep};\bm{p}^\prime,\bm{p})},
\end{align}
where $C_\mathrm{2}(t,t_\mathrm{sep};\bm{p}^\prime,\bm{p})\propto e^{-E_N(\bm{p})(t-t_\mathrm{src})}e^{-E_N(\bm{p}^\prime)(t_\mathrm{snk}-t)}$ is given by an appropriate combination of the nucleon two-point functions~\cite{Hagler:2003jd,Sasaki:2007gw}.

In the case that $\bm{p}=\bm{p}^\prime=\bm{0}$, the bare value~\footnote[2]{We recall that the renormalization factor is required to obtained the renormalized charge.
The detailed discussion of renormalization constants are given in Ref.~\cite{Tsuji:2022ric}.} of the axial-vector charge $g_A=F_A(0)$ can be evaluated for the case of  $\alpha=k$ by 
\begin{align}
\mathcal{R}^{5k}_{A_k}(t,t_\mathrm{sep};\bm{0})
=\frac{C^{5z}_{A_k}(t,t_\mathrm{sep};\bm{0},\bm{0})}{C_\mathrm{2}(t,t_\mathrm{sep};\bm{0},\bm{0})},
\end{align}
where the numerator reduces to a two-point function of the nucleon with $\bm{p}=\bm{0}$ as $C_\mathrm{2}(t,t_\mathrm{sep};\bm{0},\bm{0})=C_N(t_\mathrm{sep};\bm{0})$, 
If we could sufficiently separate the time slices as $t_\mathrm{src}\ll t\ll t_\mathrm{snk}$,
the bare axial-vector charge is read off
from $t$-independence of 
$\mathcal{R}^{5k}_{A_k}(t,t_\mathrm{sep};\bm{0})$ 
(regarded as
a plateau as a function of the operator insertion time $t$).
In practice, it is important to verify whether the plateau remains the same when $t_\mathrm{sep}$ is changed. This is called the standard plateau method.
For this method to be successful, the ground state must dominate the two-point and three-point functions of the nucleon. Thus, appropriate smearing of the nucleon interpolating operator is essential. We use  the exponentially smeared quark fields with the Coulomb gauge fixing are used for the construction of the nucleon interpolating operator. See Refs.~\cite{Shintani:2018ozy,Tsuji:2022ric,Tsuji:2023llh} for the
further details.

In the continuum, the local axial-vector current $A_\alpha(x)$ and the pseudoscalar current $P(x)$ satisfy the following operator identity,
also known as the PCAC relation:
\begin{align}
\partial_\alpha A_\alpha(x) = 2 m_\mathrm{PCAC} P(x),
\label{eq:PCAC_relation}
\end{align}
which provides a definition of the PCAC quark mass
for degenerate up and down quarks in the case of the exact SU(2) isospin symmetry.
This operator identity can be inherited in the axial Ward--Takahashi identities in terms of correlation functions.

In this study, two approaches are adopted to determine the bare value of the PCAC quark mass. One is the quantity $m_\mathrm{PCAC}^\mathrm{pion}$ evaluated from the pion two-point functions with the PCAC relation. 
The other is defined to verify the PCAC relation using the nucleon~\footnote[3]{
In Ref.~\cite{Sasaki:2007gw}, an alternative quark mass
was proposed by using the generalized Goldberger-Treiman (GGT) relation
\begin{align}
2M_N F_A(q^2)-q^2F_P(q^2)=2m_\mathrm{GGT}G_P(q^2),
\end{align}
which is satisfied among
the three nucleon form factors: the axial-vector ($F_A$),
induced pseudoscalar ($F_P$) and pseudoscalar ($G_P$) from
factors. The PCAC relation~(\ref{eq:PCAC_relation}) is essential to
leading the GGT relation. However, both the $F_P$ and $G_P$ 
form factors are strongly affected by $\pi N$-state contributions. After removing $\pi N$-state contamination~\cite{Aoki:2025taf}, 
$m_\mathrm{GGT}=m_\mathrm{PCAC}^\mathrm{pion}$ has been eventually confirmed in Ref.~\cite{Tsuji:2025quu}.
However, Eq.~(\ref{eq:mpcac_nucl}) is satisfied without isolating the ground state contribution from the excited-state contributions. 
} and is given by the nucleon three-point functions~\cite{Bali:2018qus} as follows
\begin{align}
m_\mathrm{PCAC}^\mathrm{nucl}\equiv
\frac{Z_A\partial_\alpha C_{A_\alpha}^{5k}(t,t_{\mathrm{sep}};\bm{p}^\prime,\bm{p})}{2C^{5k}_P(t,t_{\mathrm{sep}};\bm{p}^\prime,\bm{p})},
\label{eq:mpcac_nucl}
\end{align}
which requires non-zero momentum transfer $\bm{q}=\bm{p}^\prime-\bm{p}\neq \bm{0}$.
In the continuum limit, these two quantities, $m_\mathrm{PCAC}^\mathrm{pion}$,
and $m_\mathrm{PCAC}^\mathrm{nucl}$ should be identical.
Nevertheless, discrepancies between these two values may be observed on the lattice, up to terms of order $O(a)$, due to the lattice discretization uncertainties.

In this study, {\it the local axial-vector current}
is used to compute the nucleon matrix elements, which may receive a larger discretization uncertainty, instead of using an $O(a)$-improved 
current of $A_\alpha^{\mathrm{imp}}=A_\alpha+ac_A\partial_\alpha P$.
This implies that the definitions of both $m_\mathrm{PCAC}^\mathrm{pion}$,
and $m_\mathrm{PCAC}^\mathrm{nucl}$ does not take into account $O(a)$-improvement of the axial-vector current. Indeed the second term in 
the $O(a)$-improvement axial-vector current yields the $O(a)$ correction, which is given by a positive constant shift for the ground-state contribution on the value of $m_\mathrm{PCAC}^\mathrm{pion}$ as
\begin{align}
(m_\mathrm{PCAC}^\mathrm{pion})^\mathrm{imp}\equiv m_\mathrm{PCAC}^\mathrm{pion}+\frac{aZ_Ac_Am_\pi^2}{2},
\end{align}
where the correction term is proportional to $m_\pi^2$, 
since the pion two-point functions are projected 
onto zero three momentum. 

On the other hand, $m_\mathrm{PCAC}^\mathrm{nucl}$ can be computed only with finite three momentum $\bm{q}$. 
Therefore, the $O(a)$ correction yields momentum dependent term for the ground-state contribution as
\begin{align}
(m_\mathrm{PCAC}^\mathrm{nucl})^\mathrm{imp}\equiv m_\mathrm{PCAC}^\mathrm{nucl}-\frac{aZ_Ac_A q^2}{2},
\end{align}
where the correction term is proportional to the square of four-momentum transfer ($q^2$)~\footnote[4]{
In this study, we only consider the rest frame of the final state ($\bm{p}^\prime=\bm{0}$). This leads to the kinematics of $\bm{q}=\bm{p}$ and $q^2=2M_N(E_N(\bm{p})-M_N)$.
} with a negative sign.

From these considerations, the following two points can be summarized:
1) The relative signs of the contributions from the $O(a)$ correction terms in $(m_\mathrm{PCAC}^\mathrm{pion})^\mathrm{imp}$ and $(m_\mathrm{PCAC}^\mathrm{nucl})^\mathrm{imp}$ are opposite.
2) The $O(a)$ correction term in $(m_\mathrm{PCAC}^\mathrm{nucl})^\mathrm{imp}$ has explicit momentum dependence.
Consequently, a comparison between $m_\mathrm{PCAC}^\mathrm{nucl}$
and $m_\mathrm{PCAC}^\mathrm{pion}$ can elucidate the potential effects on usage of the local axial-vector current.
Conversely, if the measured values of  $m_\mathrm{PCAC}^\mathrm{pion}$ and $m_\mathrm{PCAC}^\mathrm{nucl}$ agree regardless of $q^2$, then it can be deduced that $c_A$ is 
sufficiently small to ensure that the $O(a)$ correction to the axial-vector current is negligible at low $q^2$.

\section{Numerical results}

We use the superfine ensemble of the PACS10 gauge configurations, which has been generated by the PACS Collaboration with the six stout-smeared $O(a)$-improved Wilson-clover quark action and Iwasaki gauge action 
at $\beta=2.20$. The detailed simulation parameters for all three sets of the PACS10 gauge configurations can be found in Table~\ref{tab:simulation_details}.
In the calculation of nucleon two-point and three-point functions,
the all-mode-averaging technique~\cite{Blum:2012uh,Shintani:2014vja} is used to significantly reduce the computational cost of multiple measurements and to achieve a much higher statistical accuracy.
The renormalization factor $Z_A$ is
determined by the Schr\"odinger functional method (see Appendix E in Ref.~\cite{Tsuji:2022ric}).

%%%% Axial-vector charge %%%%

%
% FIG. 1
%
%
\begin{figure}[h]
\includegraphics[width=.49\linewidth,trim=0 0 0 0,clip]{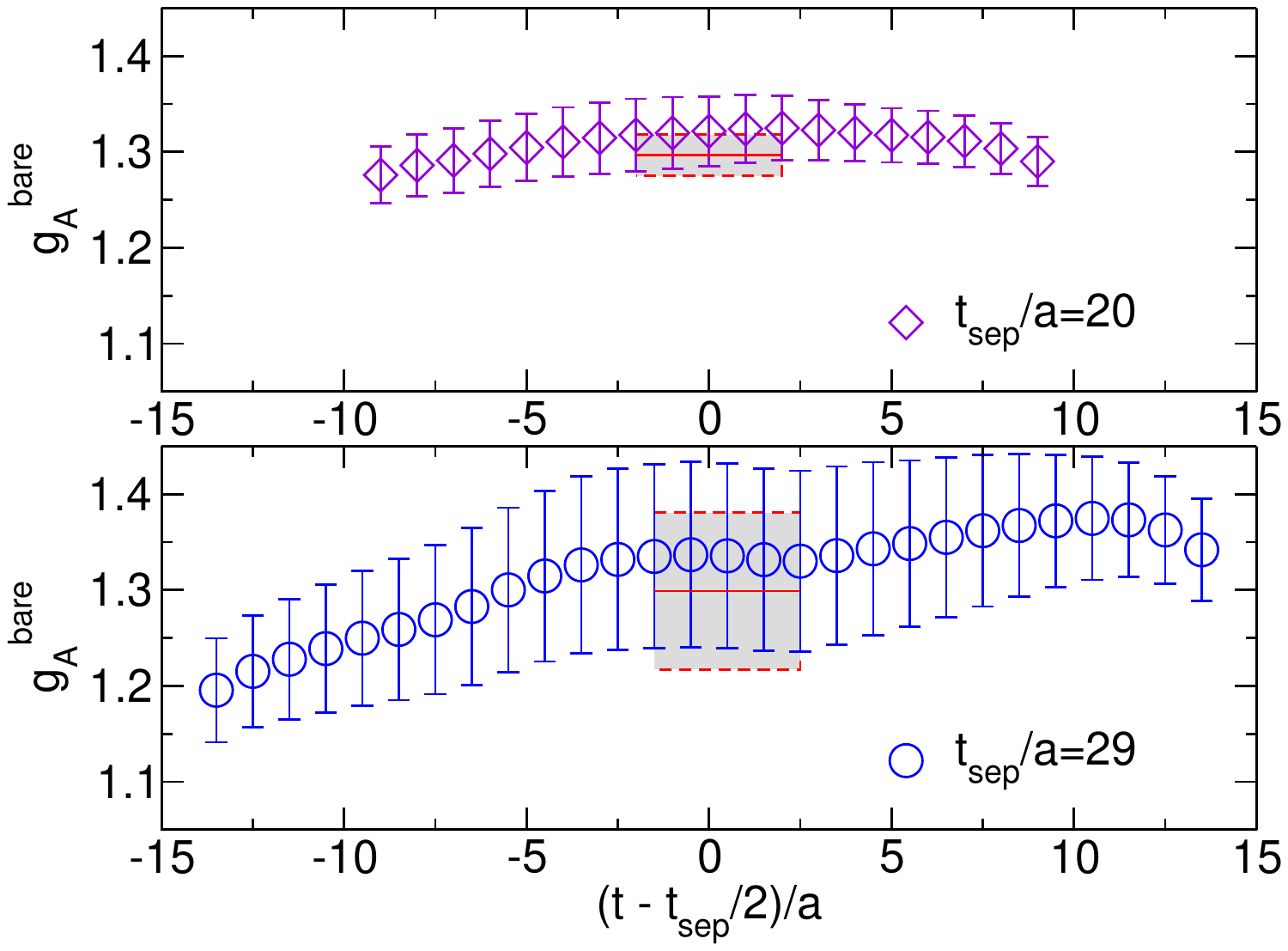}
\includegraphics[width=.49\linewidth,trim=0 0 50 0,clip]
{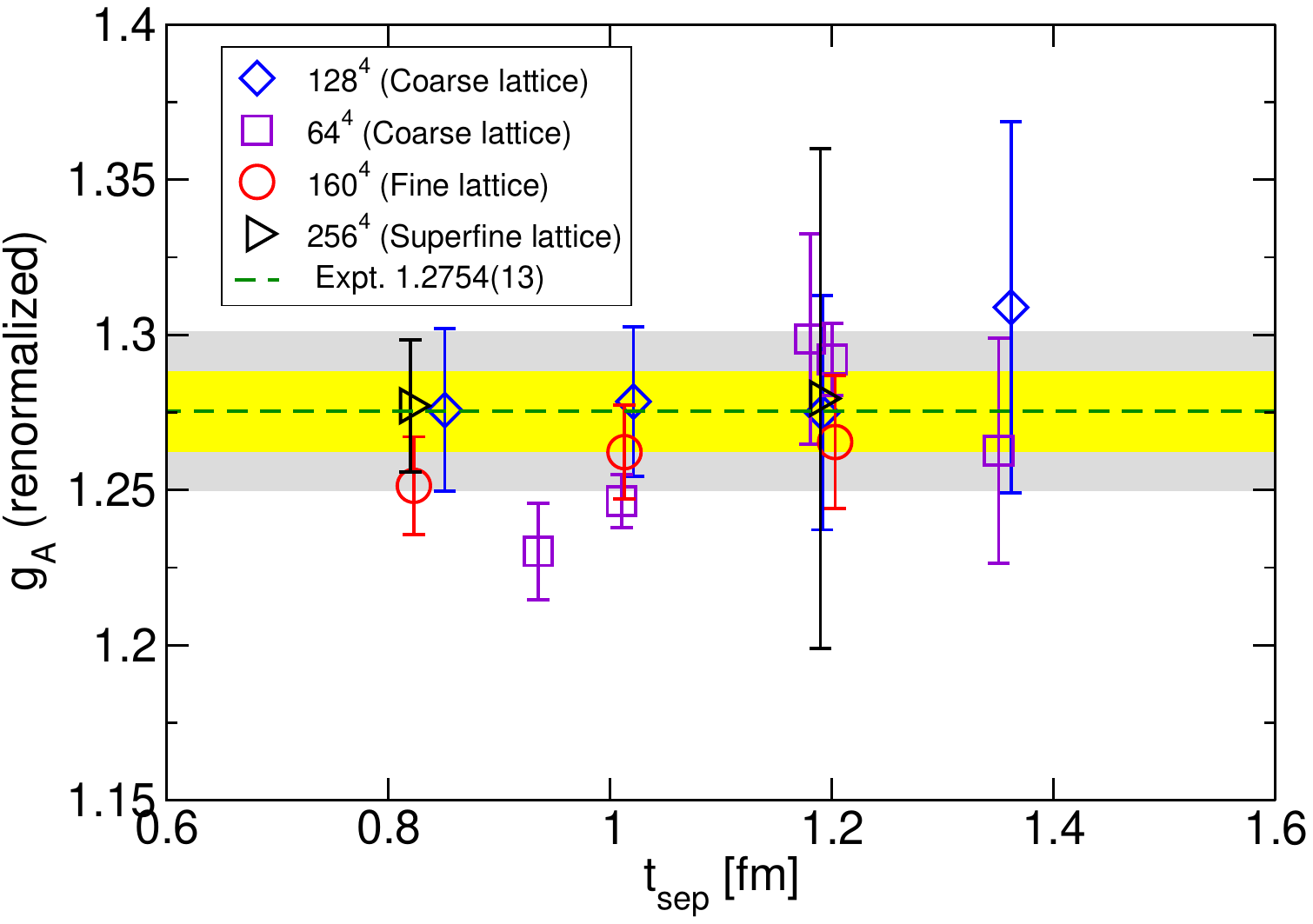}
%{./plot_ga_tsdep_p-n_XXXc_exp_ts29Jul25.pdf}
\caption{
\textbf{Left:}
The bare value of the nucleon axial-vector charge $g_A^\mathrm{bare}$ as functions of the current insertion time slice $t$ for $t_\mathrm{sep}/a=20$ and 29 in lattice units.
\textbf{Right:}
The renormalized nucleon axial-vector charge ($g_A$) as functions of $t_\mathrm{sep}$ in the physical unit.
}
\label{fig:ga_tsdep}
\end{figure}

In the left panel of Fig.~\ref{fig:ga_tsdep},
the bare value of the nucleon axial-vector charge $g_A^\mathrm{bare}$ is shown as functions of insertion time $t$ for $t_\mathrm{sep}/a=20$ and 29.
These values are obtained as the average of three-point functions
with the projection operator $\mathcal{P}_{5k}$ in all the three spatial directions $k=1,2,3$ to increase statistics.
It is observed that the plateau is formed in both cases.
Especially, the solid plateau appears with good precision for $t_\mathrm{sep}/a=20$ thanks to the elaborate tuning
of the nucleon interpolating operator using exponentially smeared 
source (sink) with Coulomb gauge fixing. Although the plateau can be seen for $t_\mathrm{sep}/a=29$, the errors are relatively large because there are insufficient measurements for satisfactory results.

The value of $g_A^\mathrm{bare}$ can be determined by the correlated constant fit using five data points in the central range of $t/a$.
In each panel, the solid line shows the fitted value, while the gray shaded band displays the fit range and 1 standard deviation for the correlated constant fit. The fit results of $g_A^\mathrm{bare}$
obtained by the two sets of $t_\mathrm{sep}$ are in agreement within the statistical error, indicating that the excited-state contamination is well under control by our choice of smearing parameters.

The results for the renormalized value of the nucleon axial-vector charge, $g_A=Z_A g_A^\mathrm{bare}$, calculated from the PACS10 superfine ensemble, along with previous results, are displayed in the right panel of Fig.~\ref{fig:ga_tsdep}. 
This provides a summary of the current status.
The horizontal dashed line shows the experimental value, while yellow and gray bands display 1\% and 2\% deviations from the experimental value.
The results of $g_A$ at the superfine lattice spacing indicate that the relative error size for $t_\text{sep}/a=20$ is comparable with that of other cases (< 2-3\%), while the relative error size for $t_\text{sep}/a=29$ remains substantially large ($\sim$ 5\%). 
Both results agree well with the experimental value and the previous
results given at the coarser lattice spacings. Therefore, the results on the superfine lattice confirm the small systematic
error associated with the finite lattice spacing effect on $g_A$, as pointed out in Ref.~\cite{Tsuji:2023llh}.

%%%% PCAC quark mass %%%%
%
% FIG. 2
%
%\begin{figure*}[h]
\begin{figure}[t]
\begin{center}
\includegraphics[width=.78\linewidth]{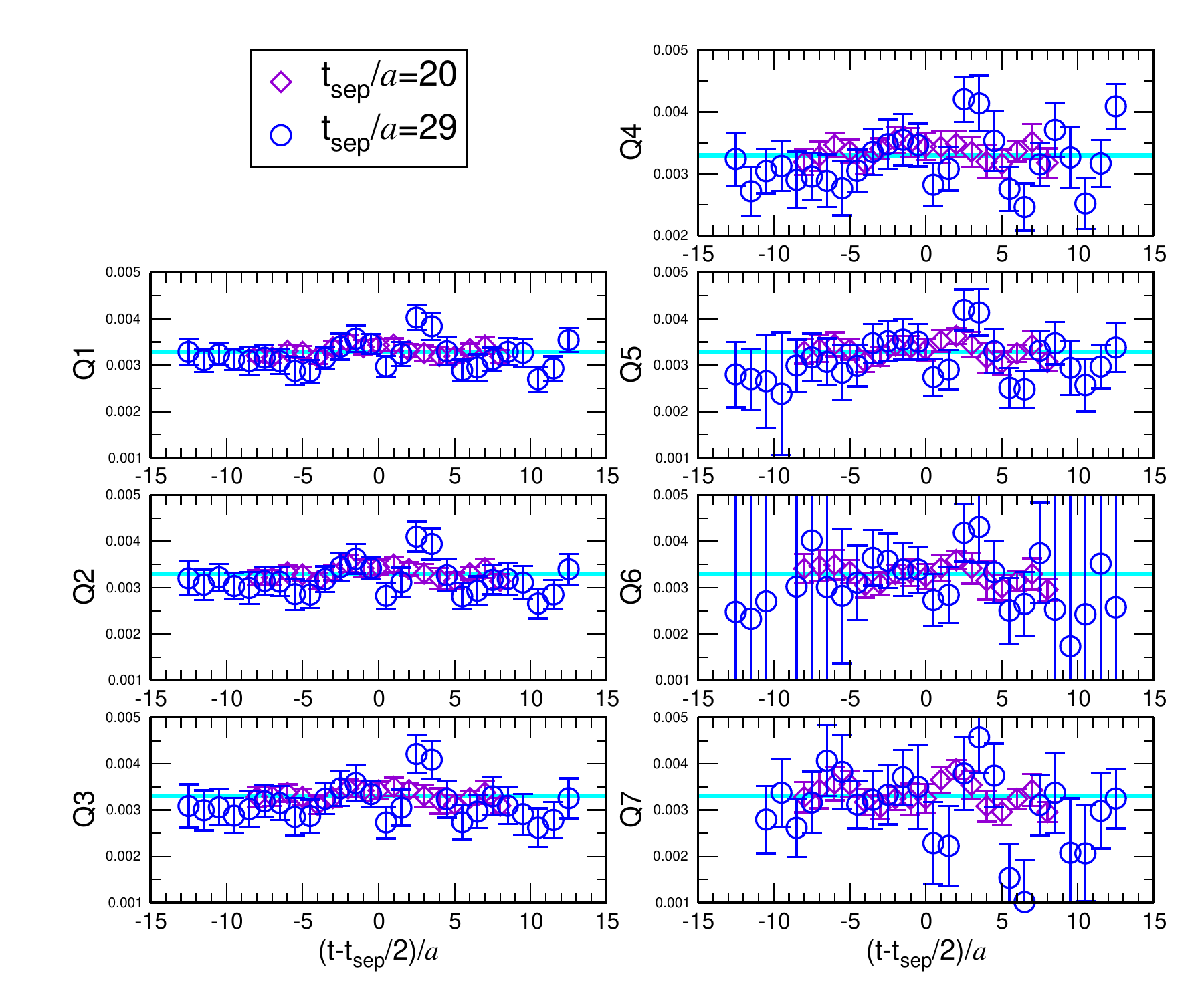}
\end{center}
\caption{
The values of $m_\mathrm{PCAC}^\mathrm{nucl}$ 
computed with $t_\mathrm{sep}/a=20$ (right triangle) and 29 (left triangle)
for all momentum transfers as functions of the current insertion time $t$. 
In each panel, the value of $m_\mathrm{PCAC}^\mathrm{pion}$ is indicated by the horizontal band.}
\label{fig:mpcac_tdep}
\end{figure}
%\end{figure*}
%

Next, we verify the PCAC relation using the nucleon three-point functions, in order to be compared to the value of $m_\mathrm{PCAC}^\mathrm{pion}$ given by the pion two-point functions. 
In Fig.~\ref{fig:mpcac_tdep}, the ratios defined in Eq.~(\ref{eq:mpcac_nucl}) are displayed for all seven finite momentum
transfers $\bm{q}$ labeled from Q1 to Q7 with two variations of $t_\mathrm{sep}/a=\{20,29\}$.
For any momentum transfer, it is observed that $m_\mathrm{PCAC}^\mathrm{nucl}$ coincides with $m_\mathrm{PCAC}^\mathrm{pion}$ within the error bars.
In particular, the consistency for $t_\mathrm{sep}/a=20$ is highly solid for all momenta in the wide range between the source and sink points.
For $t_\mathrm{sep}/a=29$, although there are still large statistical uncertainties in the high-momentum regions,
we can observe consistency at least up to Q5.

\begin{figure}[t]
\begin{center}
\includegraphics[width=.78\linewidth]{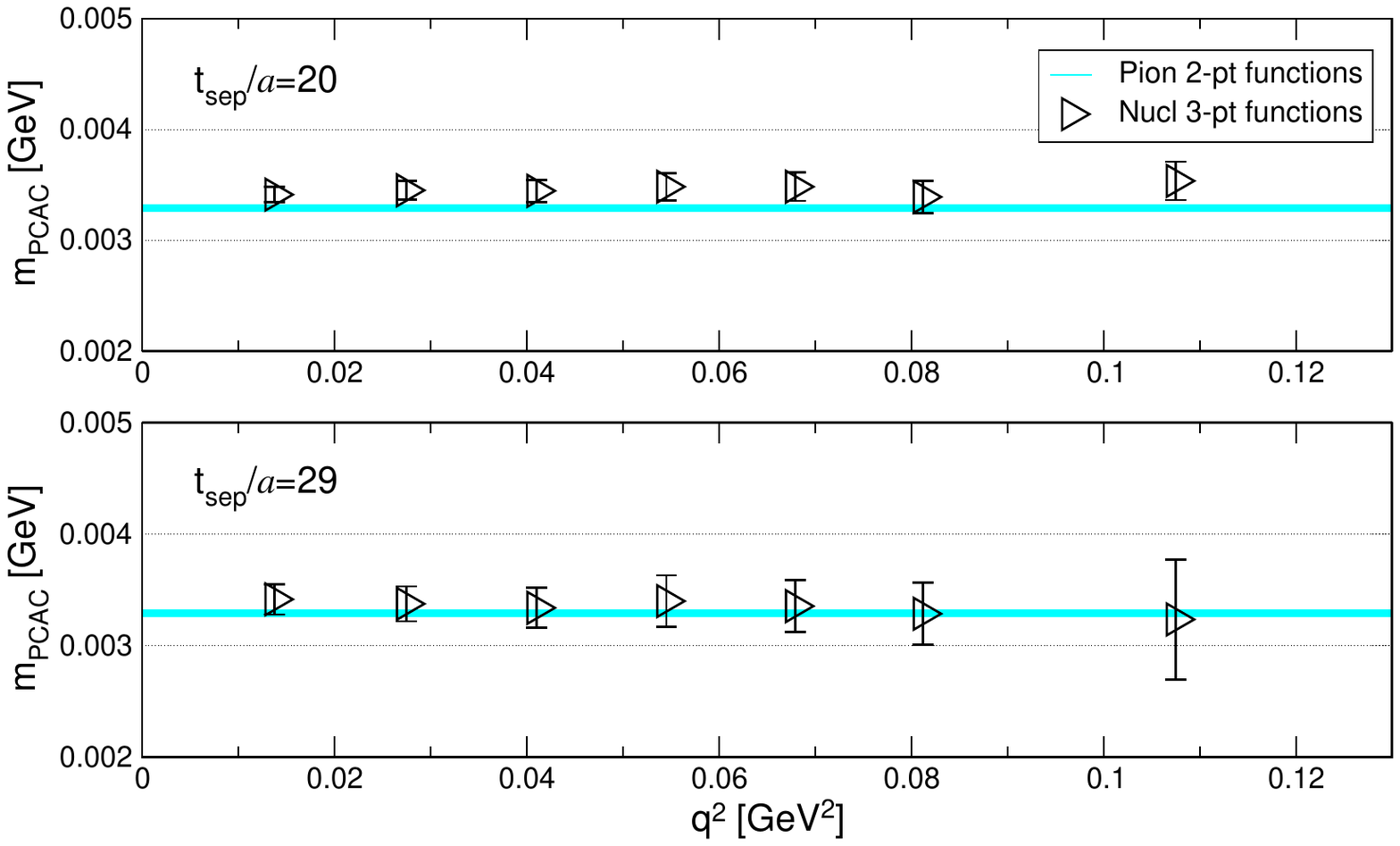}
\end{center}
\caption{
Comparison between $m_{\rm PCAC}^{\rm pion}$ (horizontal band) and $m_{\rm PCAC}^{\rm nucl}$ (open symbols) in each panel. Results for $t_{\rm sep}/a=\{20, 29\}$ are plotted from top to bottom.
}
\label{fig:mpcac_qdep}
\end{figure}

For each $q^2$ and $t_\mathrm{sep}$, we evaluate the value of $m_\mathrm{PCAC}^\mathrm{nucl}$ by weighted average using five data points in the central range of $t/a$.
Figure~\ref{fig:mpcac_qdep} shows a direct comparison of 
$m_{\mathrm{PCAC}}^{\mathrm{pion}}$ (denoted as horizontal line)
and $m_{\mathrm{PCAC}}^{\mathrm{nucl}}$ (denoted as right-triangle symbols) in the case of $t_\mathrm{sep}/a=20$ (top panel) and
29 (lower panel). All seven data points in $m_{\mathrm{PCAC}}^{\mathrm{nucl}}$ do not show strong $q^2$ dependence and accurately reproduce the value of $m_{\mathrm{PCAC}}^{\mathrm{pion}}$.
This observation indicates that even though both $m_\mathrm{PCAC}^\mathrm{pion}$ and $m_\mathrm{PCAC}^\mathrm{nucl}$ are determined from the unimproved axial-vector current, 
%the discretization errors 
the finite lattice spacing effect
in each are negligible.

%
% FIG. 3
%
%\begin{figure*}[h]
%\end{figure*}
%

Therefore, we can infer that the improvement factor, $c_A$, is very small. The observed effect is attributed to the consequence of the six stout-smeared $O(a)$-improved Wilson-clover quark action that was utilized in the generation of the PACS10 gauge configurations. In Ref.~\cite{Taniguchi:2012gew}, it was reported that 
the critical hopping parameter $\kappa_c$ and 
the non-perturbative value of the clover coefficient
$c_\mathrm{SW}$  approaches to the tree level values as the number of smearing $N_\mathrm{smr}$ increases up to $N_\mathrm{smr}=6$. This is a consequence of the smearing procedure, which helps restore the good chiral properties in the continuum theory. Indeed, the value of $c_A$ was found to be very small and consistent with zero within the statistical error~\cite{Taniguchi:2012gew}. 

In this work, it was assumed that the unimproved axial-vector current would be adequate to calculate the axial-vector matrix elements of the nucleon without any substantial effect due to the finite lattice spacing, thereby maintaining low-energy properties in continuum physics. 
This can be verified by the agreement between $m_{\rm PCAC}^{\rm pion}$ and $m_{\rm PCAC}^{\rm nucl}$.

%%%%%%%%%%%%%%%%%%%%%%%%%%%%%%
\section{Summary}
We have calculated the nucleon axial-vector charge $g_A$ and the PCAC quark masses using the third PACS10 ensemble (superfine) that is one of three sets of 2+1 flavor lattice QCD configurations
generated at the physical point on a $(10~{\mathrm{fm}})^4$ volume.
The PACS10 gauge configurations are generated with the six stout-smeared $O(a)$-improved Wilson-clover quark action and Iwasaki gauge action. Six sweeps of stout-link smearing serve to suppress ultraviolet fluctuations, thereby effectively restoring  low-energy properties in continuum physics.

The nucleon axial-vector charge $g_A$ is calculated on the superfine $256^4$ lattice. We investigate
the effects of the excited-state contamination with two sets of the source-sink separation, $t_\mathrm{sep}/a=\{20,29\}$ ($t_\mathrm{sep}\approx 0.8$ and 1.2 fm). 
The ratio of $\mathcal{R}^{5k}_{A_k}(t,t_\mathrm{sep};\bm{0})$ corresponding to the bare value of $g_A$ 
shows a clear plateau and no $t_\mathrm{sep}$ dependence,  
thanks to the elaborate tuning of the smearing parameters for the nucleon interpolating operator.
The results of $g_A$ on the superfine lattice are in agreement
with the experimental value, with relative errors of 
about 2\% for $t_\text{sep}/a=20$ and about 5\% for $t_\text{sep}/a=29$.

The PCAC quark masses calculated from the pion two-point functions ($m_\mathrm{PCAC}^\mathrm{pion}$) and the nucleon three-point functions ($m_\mathrm{PCAC}^\mathrm{nucl}$) are fairly consistent, regardless of the momentum transfer and $t_\mathrm{sep}$ 
for the definition of $m_\mathrm{PCAC}^\mathrm{nucl}$. 
This observation suggests that the effect of the $O(a)$ improvement of the axial-vector current is negligible in our calculations, since the value of $c_A$ is likely to be nearly zero. 

Further calculations with the PACS10 superfine lattice are currently underway to achieve a comprehensive study of the discretization uncertainties and subsequently take the continuum limit of target quantities.

%\clearpage
%--- acknowledgments ------------------------------------------------  
\begin{acknowledgments}
M.~N. is supported by Graduate Program on Physics for the Universe (GP-PU) of Tohoku University and JSPS Research Fellows (No. 24KJ0412). We would like to thank members of the PACS Collaboration for useful discussions. Numerical calculations in this work were performed on Oakforest-PACS in Joint Center for Advanced High Performance Computing (JCAHPC) and Cygnus  and Pegasus in Center for Computational Sciences at University of Tsukuba under Multidisciplinary Cooperative Research Program of Center for Computational Sciences, University of Tsukuba, and Wisteria/BDEC-01 in the Information Technology Center, the University of Tokyo. 
This research also used computational resources of the K computer (Project ID: hp180126)
and the Supercomputer Fugaku (Project ID: hp200188, hp210088, hp230007, hp230199, hp240028, hp240207, hp250037, and hp250218) provided by RIKEN Center for Computational Science (R-CCS),
as well as Oakforest-PACS (Project ID: hp170022, hp180051, hp180072, hp190025, hp190081,and hp200062),
Wisteria/BDEC-01 Odyssey (Project ID: hp220050) provided by the Information Technology Center of the University of Tokyo/JCAHPC.
This work is supported by the Japan Lattice Data Grid (JLDG) constructed over the SINET6 of NII.
This work was also supported in part by Grants-in-Aid for Scientific Research from the Ministry of Education, Culture, Sports, Science and Technology (Nos. 22K03612, 23H01195, 23K03428, and 25KJ0404)
and MEXT as ``Program for Promoting Researches on the Supercomputer Fugaku'' (Search for physics beyond the standard model using large-scale lattice QCD simulation
and development of AI technology toward next-generation lattice QCD; Grant Number JPMXP1020230409).

\end{acknowledgments}

% \begin{thebibliography}{99}
% \end{thebibliography}

%\clearpage
\bibliographystyle{JHEP}
\bibliography{ref}

\end{document}